\documentclass[letterpaper,twocolumn,amsmath,preprintnumbers,aps,prb,superscriptaddress]{revtex4-1}

\usepackage{graphicx}
\usepackage{bm}
\usepackage{epsfig}
\usepackage{amssymb}
\usepackage{amsfonts}
\usepackage{braket}
\usepackage{color}
\usepackage[caption=false]{subfig}

\newcommand{\la}{\langle}
\newcommand{\ra}{\rangle}

\begin{document} 
\date{\today}
\title{RKKY Interaction in Disordered Graphene}
\author{Hyunyong Lee}
\affiliation{Division of Advanced Materials Science, Pohang University
  of Science and Technology (POSTECH), Pohang 790-784, South Korea}
\author{Junghoon Kim}
\affiliation{Department of Physics, Sungkyunkwan University, Suwon
  440-746, Korea} 
\author{E. R. Mucciolo}
\affiliation{Department of Physics, University of Central Florida, 
Orlando, Florida 32816, USA}
\author{Georges Bouzerar}
\affiliation{Institut N\'eel, CNRS, d\'epartment MCBT, 25 avenue des
  Martyrs, BP 166, 38042 Grenoble Cedex 09, France}
\affiliation{School of Engineering and Science, Jacobs University
  Bremen, Bremen 28759, Germany} 
\author{S. Kettemann}
\email{s.kettemann@jacobs-university.de}
\affiliation{Division of Advanced Materials Science, Pohang University
  of Science and Technology (POSTECH), Pohang 790-784, South Korea}
\affiliation{School of Engineering and Science, Jacobs University
  Bremen, Bremen 28759, Germany}

\begin{abstract}
  We investigate the effects of nonmagnetic disorder on the
  Ruderman-Kittel-Kasuya-Yoshida (RKKY) interaction in graphene by
  studying numerically the Anderson model with on-site and hopping
  disorder on a honeycomb lattice at half filling. We evaluate the
  strength of the interaction as a function of the distance $R$
  between two magnetic ions, as well as their lattice positions and
  orientations. In the clean limit, we find that the strength of the
  interaction decays as $1/R^{3}$, with its sign and oscillation
  amplitude showing strong anisotropy. With increasing on-site
  disorder, the mean amplitude decreases exponentially at distances
  exceeding the elastic mean free path. At smaller distances, however,
  the oscillation amplitude increases strongly and its sign changes on
  the same sublattice for all directions but the armchair
  direction. For random hopping disorder, no sign change is
  observed. No significant changes to the geometrical average values
  of the RKKY interaction are found at small distances, while
  exponential suppression is observed at distances exceeding the
  localization length.
\end{abstract}

\maketitle


An unconventional behavior of the Ruderman-Kittel-Kasuya-Yoshida
(RKKY) interaction between magnetic impurities in undoped graphene was
recently reported.\cite{saremi,black,sherafati} Rather than the
conventional $1/R^{2}$ decay expected for two-dimensional systems,
where $R$ is the distance between the two magnetic moments, the RKKY
interaction is found to fall off as $1/R^{3}$ at the Dirac
(neutrality) point.
Furthermore, it was found that, due to particle-hole symmetry, only
ferromagnetic (antiferromagnetic) interactions are allowed when two
impurities are located on the same (different) sublattice.
\cite{saremi}

In a recent experiment, the authors of Ref.~\onlinecite{fuhrer}
measured the Kondo effect on graphene samples with a large number of
vacancies, confirming that these defects induce local magnetic
moments.\cite{peres2006,ugeda} Thus, upon increasing the control over
the location of such defects, one might be able to also measure the
RKKY interaction as a function of distance and location of local
moments. Indeed, a direct detection of the RKKY interaction is
feasible with the recent development of a technique to measure the
magnetization curves of individual atoms using spin-polarized scanning
tunneling spectroscopy.\cite{wiesendanger1,wiesendanger2} With this
technique, the orientation and distance dependence of the exchange
interactions can be observed precisely.

The influence of disorder on the RKKY interaction in conventional
metals has been thoroughly studied.\cite{chatel,zyuzin,abrahams} These
studies found that the main 
effect of weak disorder is to randomize the electron phase, resulting
in an exponential decrease of the ensemble-averaged
interaction amplitude with distance. However, the average does not
properly characterize the typical interaction strength, as any
particular disorder configuration has long-range correlations. Indeed,
the typical value, identified as the
geometrical average ($J^{\textrm{geo}}_{\textrm{RKKY}} \equiv
e^{\langle(1/2) \ln [J_{\textrm{RKKY}}]^{2}\rangle_{\textrm{avg}}}$)
is found to have the same power-law behavior with distance as the
amplitude of the interaction in the clean limit. Consequently, at
least for conventional metals, weak disorder is not likely to cause
any critical change in physical properties which derive from the RKKY
interaction. It is only when the system approaches the localized
regime that the geometrical average is exponentially
suppressed.\cite{sobota}

In light of these facts, our study focuses on two main questions. The
first is how a pair of magnetic impurities in disordered graphene will
interact in general. We consider impurities located along any lattice
orientation, and not only along the zigzag and armchair lines. The
second is how this interaction changes with increasing disorder strength.

Let us begin by considering a general expression for the RKKY exchange
coupling constant in terms of the unperturbed (disorder-free)
electronic Green's function $G^{(0)}({\bf r}_i,{\bf
  r}_j,\omega)$,\cite{white}

\begin{eqnarray}
J_{\textrm{RKKY}} & = & J^2 \frac{S(S+1)}{4 \pi S^{2}} \int d\omega \,
f(\omega)\, \textrm{Im} \left[ G^{(0)}(\bm{r}_j,\bm{r}_i, \omega)
  \right. \nonumber \\ & & \times \left.
  G^{(0)}(\bm{r}_i,\bm{r}_j,\omega)\right] \\ & = & J^2
\frac{S(S+1)}{4 \pi S^{2}} \textrm{Im} \int d\omega \, f(\omega)
\nonumber \\ & & \times \sum_{n,m} \frac{F^{ij}_{nm}}{(E_n - \omega +
  i\delta)(E_m - \omega + i\delta)}.
\label{eq:RKKY_original}
\end{eqnarray}
Here, $J$ is the local coupling constant between the localized
magnetic impurities and the itinerant electrons, $S$ is the magnitude
of the impurity spin, $i$ ($j$) is the site index of a magnetic
impurity located at position $\bm{r}_{i}$ ($\bm{r}_{j}$), $ f(\omega)
= [e^{(\omega -\mu)/T} + 1]^{-1}$ is the Fermi-Dirac distribution
function, and $F^{ij}_{nm} = \psi^*_n(\bm{r}_i) \psi_n(\bm{r}_j)
\psi^*_m(\bm{r}_j) \psi_m(\bm{r}_i)$, with $\psi_n(\bm{r}_i)$ denoting
the eigenfunction corresponding to the eigenenergy $E_n$ of the
unperturbed electronic Hamiltonian (i.e., in the absence of magnetic
disorder). The lattice constant $a$ and $\hbar$ are set to unity in
all calculations.

Using a zero-temperature approximation ($T=0$) and changing to
an integral form, Eq.~\eqref{eq:RKKY_original} can be recast as
\begin{equation}
  J_{\textrm{RKKY}} = -J^2\frac{S(S+1)}{2S^2} \int_{E<0} dE \int_{E'>0} dE'\
  \frac{F(E,E')}{E-E'},
  \label{eq:J-KPM}
\end{equation}
where $F(E,E') = \textrm{Re}[\rho_{ji}(E)\rho_{ij}(E')]$, $\mu$ is the
Fermi energy, and $\rho_{ij}(E) = \braket{i|\delta(E-H)|j}$, which can
be calculated numerically using the kernel polynomial method
(KPM).\cite{alexander} In the KPM, the matrix elements $\rho_{ij}(E)$
are expressed as sums over order-$M$ Chebyshev polynomials on the energy $E$
with coefficients obtained through an efficient recursion relation
involving matrix elements of the system Hamiltonian.
As our unperturbed electronic Hamiltonian with on-site disorder, we
employ the single-band Anderson tight-binding model on a honeycomb
lattice,
\begin{equation}
H = -t\sum_{\la i,j \ra} c_{i}^+ c_{j} + \sum_{i} w_i~c_{i}^+ c_{i},
\label{eq:Hamiltonian}
\end{equation}
where $t$ ($\approx$ 2.67 eV for graphene) is the hopping energy,
$c_i~(c_i^+)$ annihilates (creates) an electron at site $i$, $w_i$ is
the on-site random disorder energy distributed uniformly between
$[-W/2, W/2]$, and $\la i,j \ra$ denote nearest-neighbor
sites. Periodic boundary conditions are used for all calculations. For
clean systems ($W=0$), the Chebyshev polynomials are calculated up to
$M=3\times 10^{3}$ on a lattice with $5\times 10^{5}$ sites.

\begin{figure}[!ht]
  \captionsetup[subfloat]{font = {bf,up}, position = top}
  \subfloat[~~~~~~~~~~~~~~~~~~~~~~~~~~~~~~~~~~~~~~~~~~~~~~~~~~~~~~~~~~~~~~~~~~]{\includegraphics[width=0.49\textwidth]{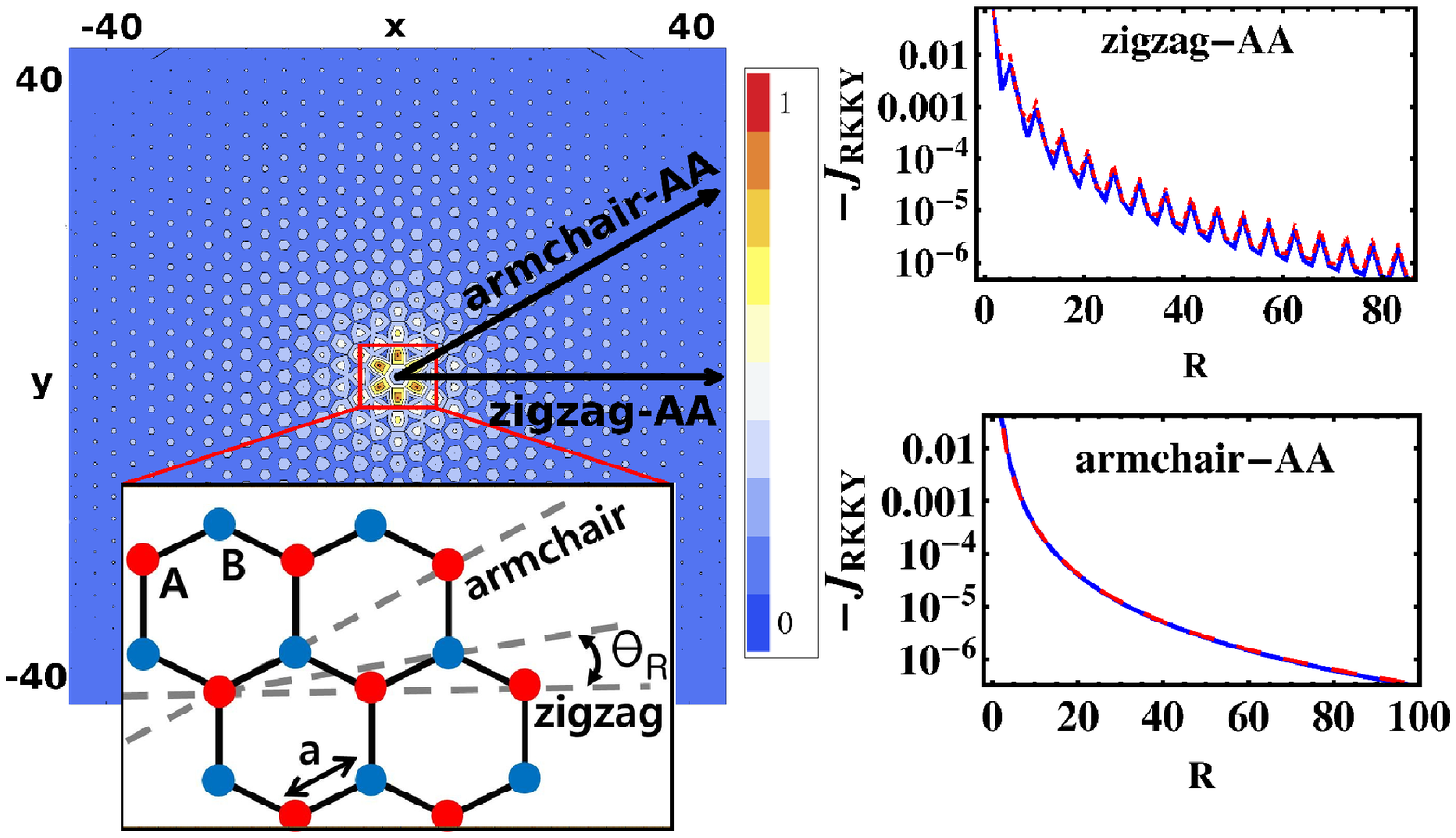}}\\
  \subfloat[~~~~~~~~~~~~~~~~~~~~~~~~~~~~~~~~~~~~~~~~~~~~~~~~~~~~~~~~~~~~~~~~~~]{\includegraphics[width=0.49\textwidth]{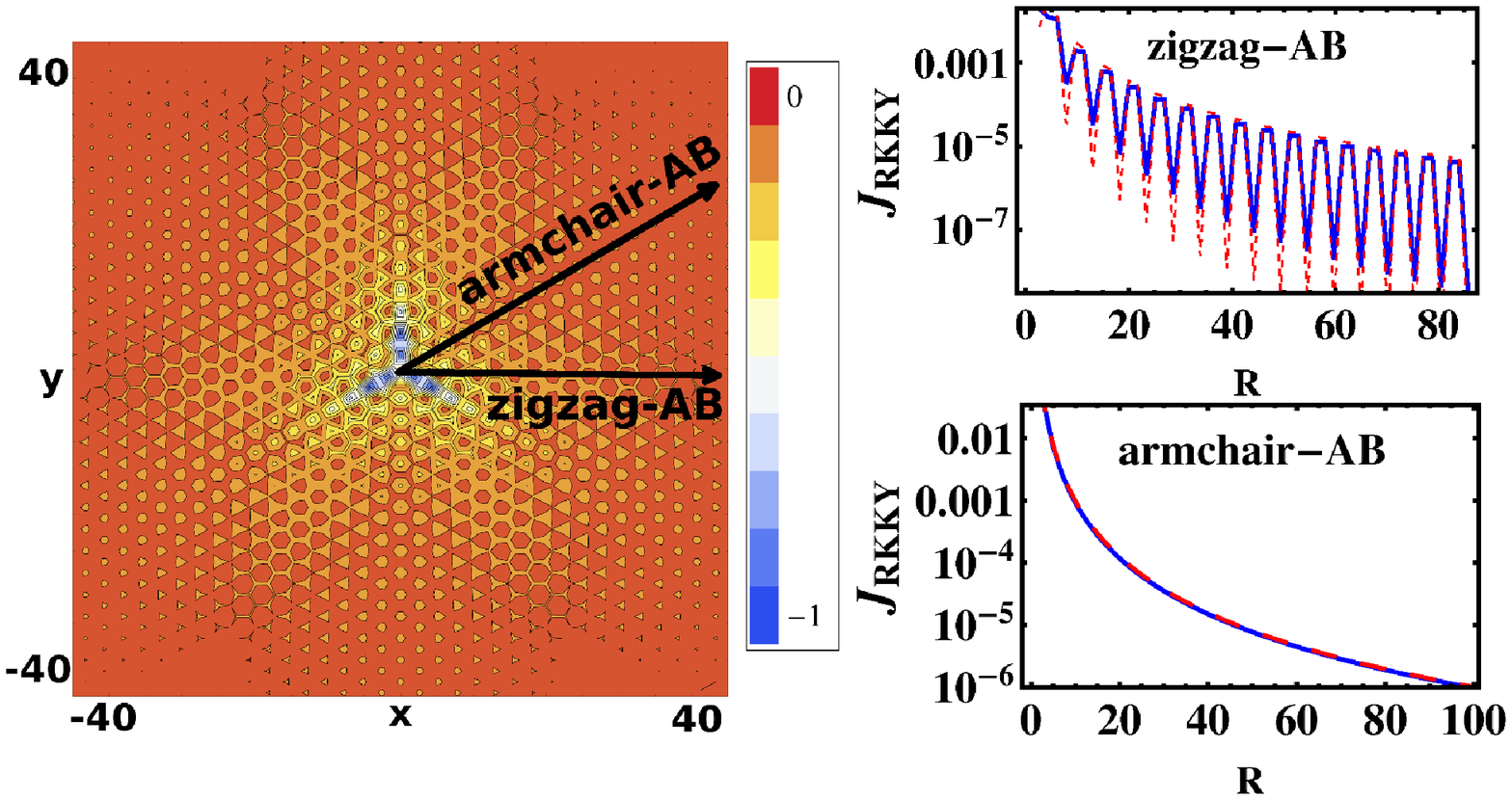}} 
 \caption{(Color online) Plots of the RKKY interaction strengths
    between a magnetic impurity at the origin and another at: (a) a
    site from the same sublattice (AA) and (b) a site from a different
    sublattice (AB). In the contour plots, the amplitudes are
    multiplied by the square of the distance to facilitate visualization. The
    lattice constant is set to unity. The numerical data is for clean
    graphene ($W = 0$). Calculations using the kernel polynomial
    method and lattice Green's function method are represented as
    solid blue and dashed red lines, respectively.}
  \label{fig:undoped-RKKY} 
\end{figure}

The RKKY interaction coupling constant between two magnetic impurities
is calculated using Eq.~\eqref{eq:J-KPM}, of which the results for the
clean limit are shown in Fig.~\ref{fig:undoped-RKKY}. In order to
better visualize the behavior of the amplitude in the contour plots,
we have multiplied $J_{\rm RKKY}$ by $R^2$, resulting in a smoother
($1/R$) decay. The interactions along the zigzag and armchair
directions are shown separately by line plots in
Fig.~\ref{fig:undoped-RKKY}. These results are in excellent agreement
with previous studies.\cite{saremi, black, sherafati} The authors of
Ref.~\onlinecite{sherafati} used a lattice Green's function method to
obtain an RKKY interaction of the form 
\begin{eqnarray}
J_{\textrm{AA}}^{0} & = & -J^2 \frac{1 + \cos[(\bm{K}^{+}-\bm{K}^{-})
    \cdot \bm{R}]}{R^3},
\label{eq:rkky_dirac_AA}
\\ J_{\textrm{AB}}^{0} & = & J^2 \frac{3 + 3\cos[(\bm{K}^{+}-\bm{K}^{-})
    \cdot \bm{R} + \pi -2\theta_R]}{R^3},
\label{eq:rkky_dirac_AB}
\end{eqnarray}
where all the coefficients are set to unity, $\bm{K}^{\pm}=(\pm
2\pi/3\sqrt{3},2\pi/3)$ are the Dirac points in the Bloch momentum
space, ${\bf R}={\bf r}_i - {\bf r}_j$, and $\theta_R$ is defined in the
inset of Fig.~\ref{fig:undoped-RKKY}a. For a direct comparison, plots
of Eqs.~\eqref{eq:rkky_dirac_AA} and \eqref{eq:rkky_dirac_AB} are
shown in Fig. \ref{fig:undoped-RKKY} along with the results calculated
from Eq.~\eqref{eq:J-KPM}. As expected from the particle-hole symmetry
of the spectrum, the magnetic impurity on the origin has ferromagnetic
correlations with the impurities on the same sublattice
(Fig.~\ref{fig:undoped-RKKY}a), while antiferromagnetic correlations
develop for impurities on different sublattices
(Fig.~\ref{fig:undoped-RKKY}b).

In order to investigate the effect of on-site nonmagnetic disorder in
graphene, we consider $1.6\times 10^{3}$ different disorder
configurations for each value of $W$ and then evaluate the matrix
elements $\rho_{ij}$ through the KPM with $M=5\times 10^{3}$ on a
lattice with $1.8\times 10^{5}$ sites.

\begin{figure}[!ht]
\includegraphics[width=0.49\textwidth]{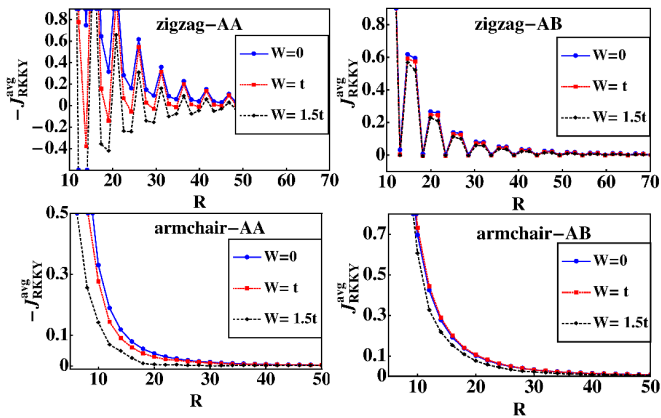}
\caption{(Color online) Plots of the RKKY interaction strength along
 the (a) zigzag and (b) armchair directions in the diffusive regime,
 as averaged over $1.6\times 10^{3}$ different disorder
 configurations. A lattice with $1.8\times 10^{5}$ sites and a polynomial
 degree cutoff of $M = 5 \times 10^{3}$ are used in these numerical
 calculations.}
\label{fig:Avg_total} 
\end{figure}

\begin{figure}[!ht]
\includegraphics[width=0.35\textwidth]{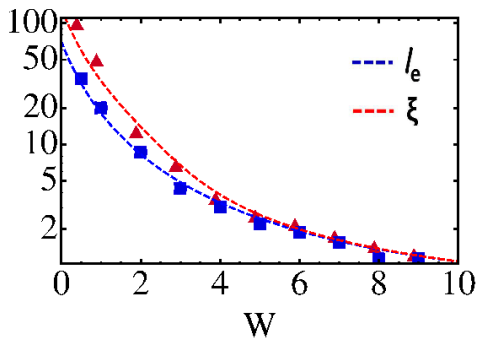}
 \caption{(Color online) Plots of the mean free path $l_{e}$ and the
localization length $\xi$ as functions of the disorder strength
 $W$ (in units of $t$). The blue dashed line represents a fitting
to the relation $l_e = c_1/W^2$ with $c_1 = 70$, whereas the red
dashed line represents the resulting localization length,
with $v_{\textrm{F}} \Lambda = 10$ (see text).}
\label{fig:length_scale}
\end{figure}

For weak (strong) disorder strength, the system is in the diffusive
(localized) regime, where the actual value of $W$ for which this
crossover occurs depends on the lattice size and has been
determined by evaluating the localization length (see
Fig.~\ref{fig:length_scale}). The average amplitude of the RKKY
interaction in the diffusive regime is shown in
Fig.~\ref{fig:Avg_total}. Similar to conventional metals, the
interaction decays exponentially with increasing disorder strength as
\begin{equation}
J^{\textrm{avg}}_{\textrm{RKKY}} \sim
J^{\textrm{clean}}_{\textrm{RKKY}}~e^{-R/l_{e}},
\label{eq:J_avg}
\end{equation}
where $l_{e}$ is the mean free path and
$J^{\textrm{clean}}_{\textrm{RKKY}}$ is the interaction amplitude in
the clean limit. It is worth noticing that the sign of the interaction
oscillates when the impurities are located along the zigzag-AA
direction.

\begin{figure}[!ht]
  \captionsetup[subfloat]{font = {bf,up}, position = top}
  \centering
  \subfloat[~~~~~~~~~~~~~~~~~~~~~~~~~~~~~~~~~~~~~~~~~~~~~~~~~~~~~~]
  {\includegraphics[width=0.4\textwidth]{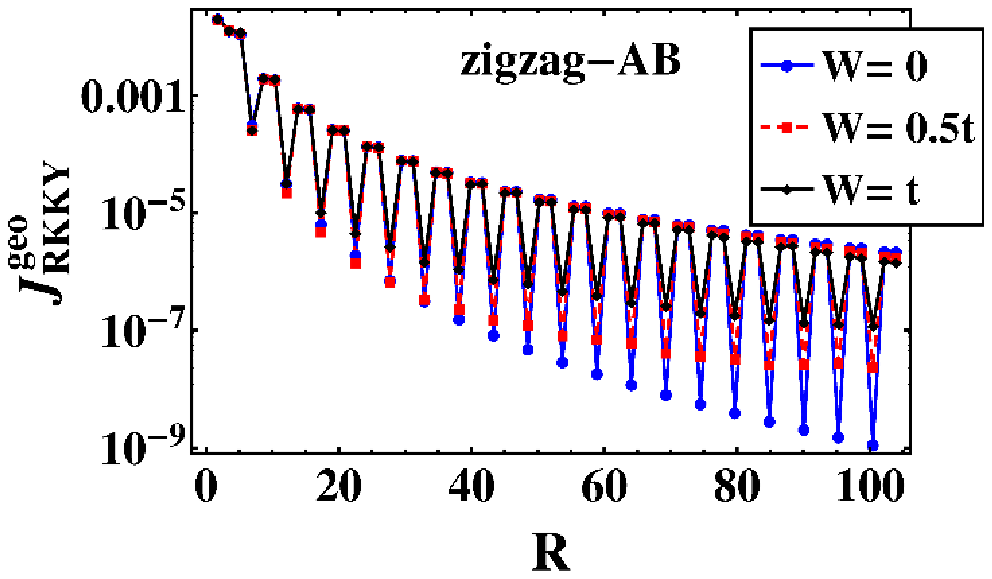}}\\
  \subfloat[~~~~~~~~~~~~~~~~~~~~~~~~~~~~~~~~~~~~~~~~~~~~~~~~~~~~~~]
  {\includegraphics[width=0.4\textwidth]{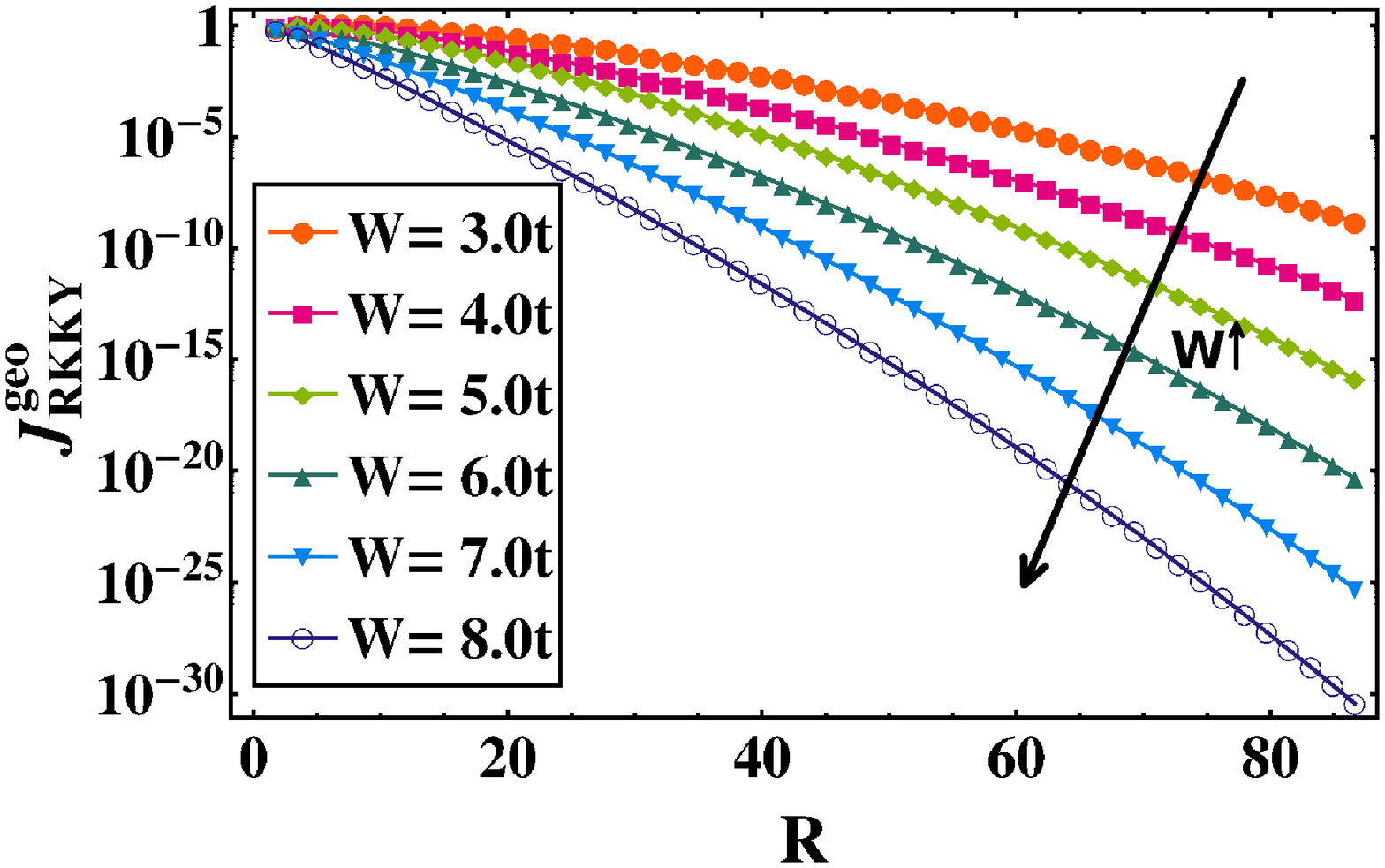}}
  \caption{(Color online) Plots of the geometrical average over
    $1.6\times 10^{3}$ different disorder configurations of the RKKY
    interactions for (a) weak and (b) strong disorder. The same
    lattice size and polynomial cutoff of Fig.~\ref{fig:Avg_total} are
    used.} 
  \label{fig:typical_value} 
\end{figure}

To better characterize the amplitude of the interaction, we have also
calculated the geometrical average
($J^{\textrm{geo}}_{\textrm{RKKY}}$) for both diffusive and localized
regimes (Fig.~\ref{fig:typical_value}). In
Fig.~\ref{fig:typical_value}a, one can see that the geometrical
average for a weakly disordered system remains long ranged and has a
decaying behavior similar to the clean system. As mentioned earlier,
studies of conventional metals \cite{chatel,zyuzin,abrahams} have
shown that the geometrical average (i.e., the typical value) of the
RKKY interaction in weakly disordered systems has a power law dependence
with the same exponent of the clean limit, namely,
$J_{RKKY}^{\textrm{geo}} \sim 1/R^{\alpha}$ (e.g., $\alpha = 2$ in a
two-dimensional electron gas). Due to the unconventional distance
dependence (Eqs.~\eqref{eq:rkky_dirac_AA} and
  \eqref{eq:rkky_dirac_AB}) caused by the pseudogap at the Dirac point
  of clean graphene, one may expect two possibilities. If the pseudogap is
not filled by disorder, the geometrical average value of the amplitude
is expected to have the same exponent of the clean system, namely,
$\alpha = 3$. However, if it is filled, then the geometrical average
value should approach the conventional power law of a two-dimensional
electron gas, namely, $\alpha = 2$.
Our calculations show that the former is the correct answer. This is
in accordance with the fact that short-range disorder preserves the
pseudogap in graphene.\cite{amini} Therefore, the presence of weak
short-range disorder in undoped graphene is not anticipated to induce
any major change in physical properties related to the RKKY
interaction.

The situation is drastically different in the localized regime, where
the geometrical average values is exponentially suppressed with
distance, as shown in Fig.~\ref{fig:typical_value}b. This behavior is
captured by the following relation,\cite{sobota}
\begin{equation}
  J_{\textrm{RKKY}}^{\textrm{geo}} \sim e^{-R/\xi},
\label{eq:J_geo}
\end{equation}
where $\xi$ is the localization length. Fig.~\ref{fig:length_scale}
presents the mean free path and the localization length obtained by
fitting the relations Eq.~\ref{eq:J_avg} and Eq.~\ref{eq:J_geo} to the
numerical data. For $W = t$, the localization length is about $10^{2}$,
which is close to the longest linear distance possible in our
calculations, namely $R_{\rm max} = 60\sqrt{3}$. Therefore, the system
crosses over from the diffusive to the localized regime around $W \sim
t$. For uncorrelated, short-range disorder, which allows for
intervalley scattering, the localization length is given by $\xi =
l_{e} \exp (\pi\sigma/\textrm{G}_0)$,\cite{lherbier,neto} where
%
$
  \sigma =\frac{4}{\pi} \Big[ \frac{(v_{\textrm{F}} \Lambda)^2}{(v_{\textrm{F}} 
    \Lambda)^2 + W^4}\Big],
$
%
$v_{\textrm{F}}$ denotes the Fermi velocity, $\Lambda$ is the energy
cutoff, and $\textrm{G}_0 = e^{2}/h$ is the conductance quantum.

 It is well known that the mean free path is inversely proportional to
the square of disorder strength ($l_e \sim 1/W^2$). Therefore, one
expects the localization length to obey the relation
$
  \xi \approx (c_1/W^2)\exp\Big[ \frac{4(v_{\textrm{F}}
      \Lambda)^2}{(v_{\textrm{F}} \Lambda)^2 + W^4}\Big],
$
where $c_1$ is a fitting constant. Indeed, these relations fit
reasonably well the numerical data, as shown in
Fig.~\ref{fig:length_scale}.

\begin{figure}[!ht]
  \captionsetup[subfloat]{font = {bf,up}, position = top}
  \centering
  \subfloat[~~~~~~~~~~~~~~~~~~~~~~~~~~~~~~]
  {\includegraphics[width=0.24\textwidth]{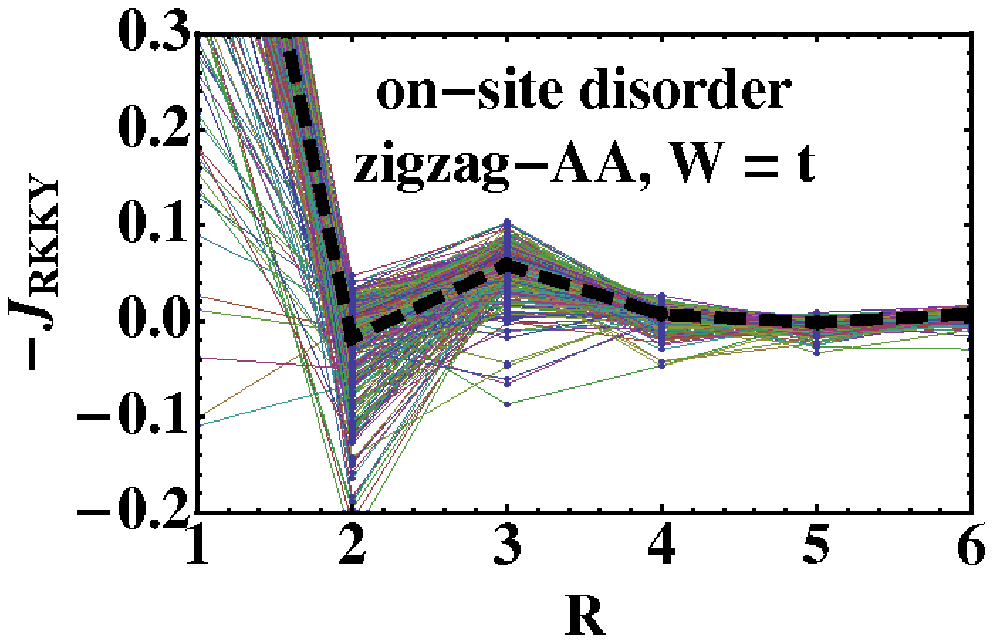}}                
  \subfloat[~~~~~~~~~~~~~~~~~~~~~~~~~~~~~~]
  {\includegraphics[width=0.24\textwidth]{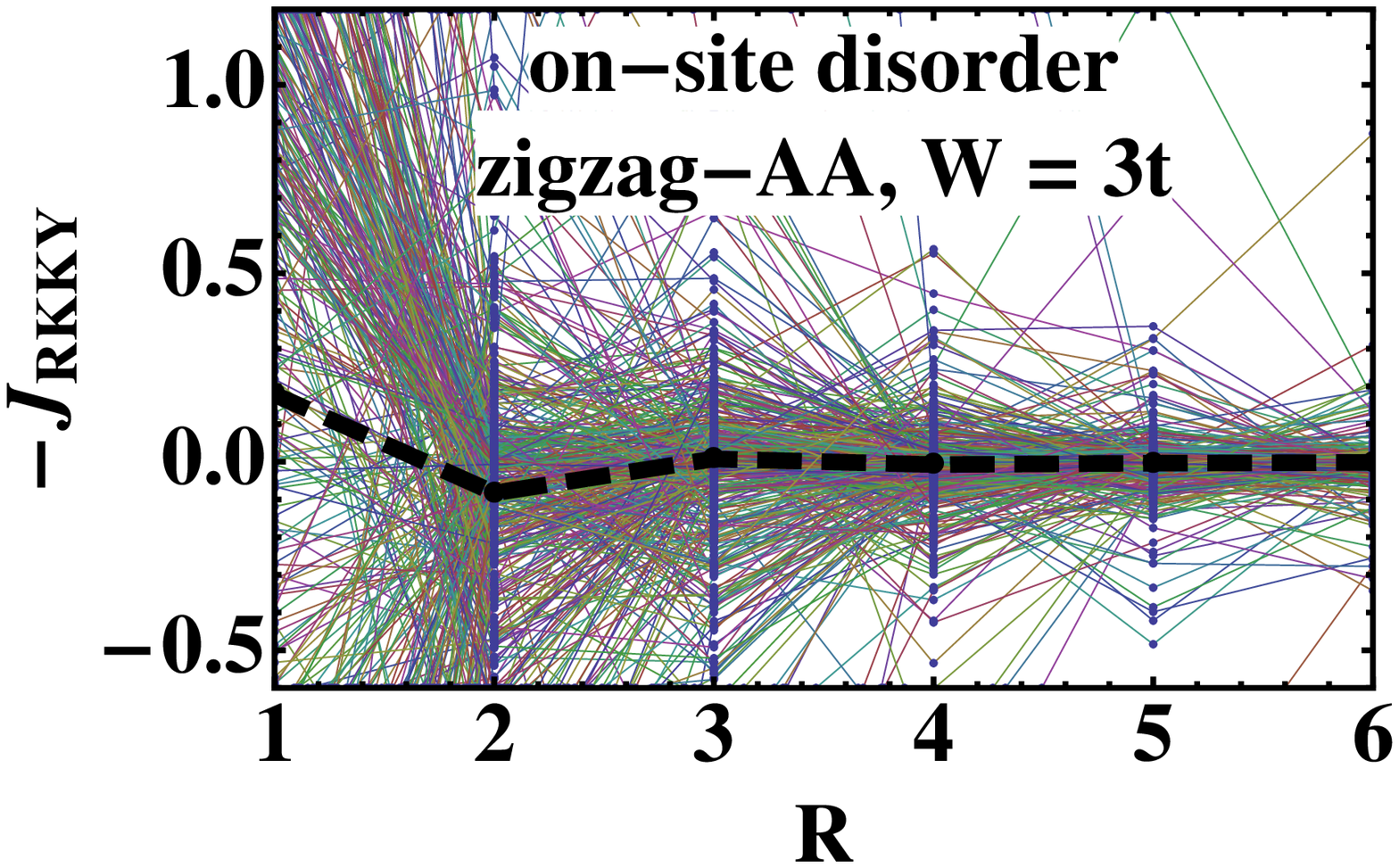}}\\ 
  \subfloat[~~~~~~~~~~~~~~~~~~~~~~~~~~~~~~]
  {\includegraphics[width=0.24\textwidth]{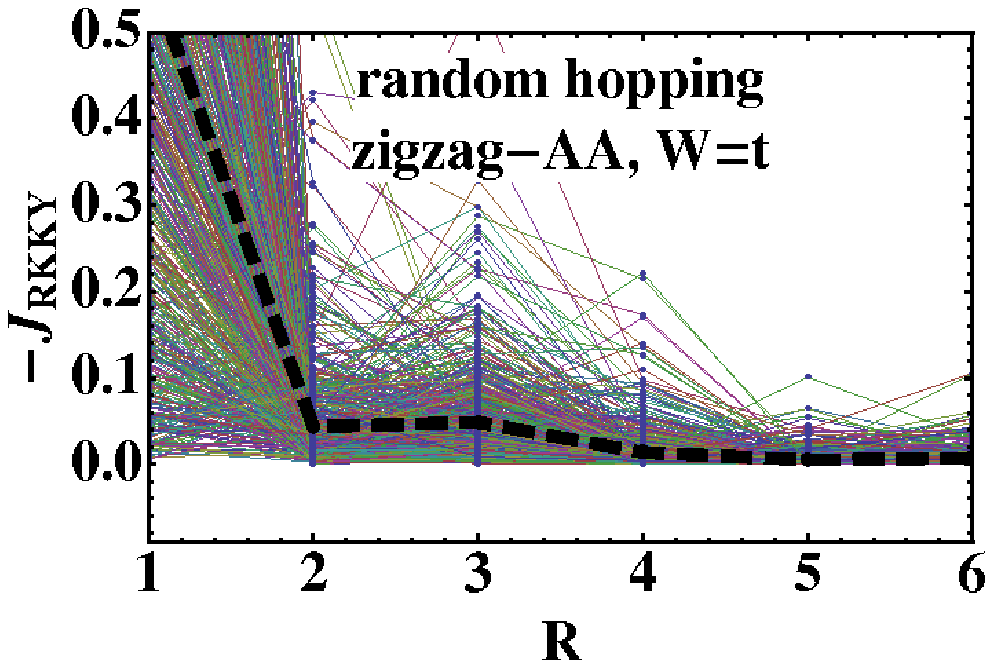}}
  \subfloat[~~~~~~~~~~~~~~~~~~~~~~~~~~~~~~]
  {\includegraphics[width=0.24\textwidth]{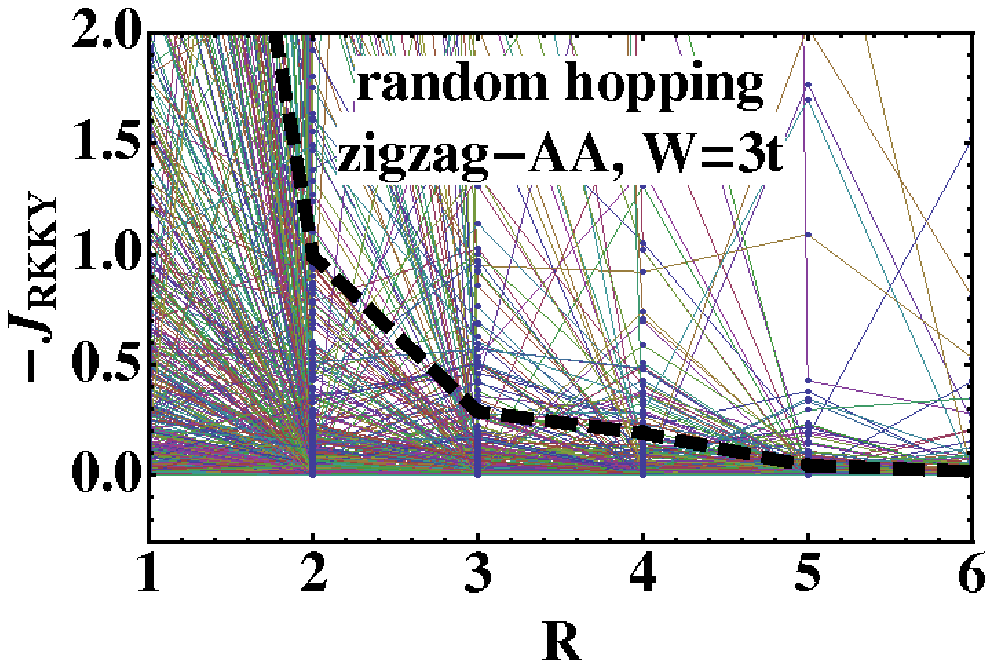}}
  \caption{ Plots of the RKKY interactions along the zigzag-AA
directions with diagonal disorder (a) $W = t$, (b) $W = 3t$ and
off-diagonal~(random hopping) (c) $W = t$, (d) $W = 3t$ for $5\times 10^{2}$
realizations. A lattice with $2\times 10^{4}$ sites and a polynomial degree
cutoff of $M = 10^{3}$ are used in these numerical calculations. The black
dashed line is the averaged interaction.}
\label{fig:random_hopping} 
\end{figure}

To find out the effect of disorder with no sublattice symmetry
breaking, we added randomness to the hopping integral and eliminated
on-site disorder ($w_i = 0$),
\begin{eqnarray}
  H = - \sum_{\la i,j \ra} t_{ij} c_{i}^+ c_{j},
\end{eqnarray}
where $t_{ij} = t + \Delta t_{ij}$, with $\Delta t_{ij}$ being
distributed uniformly between $[-W/2, W/2]$. We perform the same
calculations of the on-site disorder case, but now with a lattice of
$2\times 10^{4}$ sites and a Chebyshev polynomial cutoff $M = 10^{3}$. For
comparison, we plot the results together with those for the on-site
disorder calculations in Fig.~\ref{fig:random_hopping}. A total of
$5\times 10^{2}$ configurations of disorder are used, with the thick
dashed line indicating the average value. While the on-site disorder
generates random fluctuations in the sign and amplitude of the RKKY
interaction (Figs.~\ref{fig:random_hopping}a,b), the hopping disorders
affect only the amplitude, even in the presence of very strong
randomness~(Figs.~\ref{fig:random_hopping}c,d).

In conclusion, we have confirmed that the RKKY interactions in clean
graphene has a strong anisotropy of its sign and oscillation
amplitude, and it decays as $1/R^3$ for all directions. Increasing the
amount of nonmagnetic, on-site disorder causes the averaged amplitude
of the RKKY interaction to decrease exponentially at distances
exceeding the elastic mean free path, similarly to what is obtained
for conventional metals. At smaller distances, however, the
fluctuations of the amplitude are found to increase strongly, with
sign oscillations even for a pair of magnetic impurities located on
the same sublattice, for all directions except the armchair
direction. When the randomness is instead applied to the
hopping~(off-diagonal disorder), the sign oscillations disappear. This
shows that these sign changes at weak disorder potential are caused by
the breaking of the sublattice symmetry, since off-diagonal disorder
preserves this symmetry. Our calculations also confirm that the
geometrical average of the RKKY interaction in disordered graphene has
the same power law decay at short distances as in the clean
case. However, it is exponentially suppressed at distances exceeding
the localization length. We plan, to extend these studies by
considering the effects of long-range disorder and resonant
impurities. 

We gratefully acknowledge that this research was supported by WCU
(World Class University) program through the National Research
Foundation of Korea funded by the Ministry of Education, Science and
Technology(R31-2008-000-10059-0), Division of Advanced Materials
Science. ERM acknowledges partial support through the NSF award DMR
1006230. ERM and GB thank the WCU AMS for its hospitality.




\begin{thebibliography}{99}

\bibitem{saremi} S. Saremi, Phys. Rev. B {\bf 76}, 184430 (2007).

\bibitem{black} A. M. Black-Schaffer, Phys. Rev. B {\bf 81}, 205416
  (2010).

\bibitem{sherafati} M. Sherafati and S. Satpathy, Phys. Rev. B {\bf
  83}, 165425 (2011).

\bibitem{fuhrer} J.-H. Chen, W. G. Cullen, E. D. Williams, and
  M. S. Fuhrer, Nature Phys. {\bf 7}, 535 (2011).
 
\bibitem{peres2006} N. M. R. Peres, F. Guinea, and A. H. Castro Neto,
  Phys. Rev. B {\bf 73}, 125411 (2006).

\bibitem{ugeda} M. M. Ugeda, I. Brihuega, F. Guinea, and
  J. M. Gomez-Rodriguez, Phys. Rev. Lett. {\bf 104}, 096804 (2010).

\bibitem{wiesendanger1} F. Meier L. Zhou, J. Wiebe, and
  R. Wiesendanger, Science {\bf 320}, 82 (2008).

\bibitem{wiesendanger2} L. Zhou {\it et al.}, Nature Phys. {\bf 6},
  187 (2010).

\bibitem{chatel} P. F. de Chatel, J. Magn. Magn. Mater. {\bf 23}, 28
  (1981).

\bibitem{abrahams} A. Jagannathan, E. Abrahams, and M. J. Stephen,
  Phys. Rev. B {\bf 37}, 436 (1988).

\bibitem{zyuzin} A. I. Zyuzin and B. Z. Spivak, JEPT Lett. {\bf 43},
  243 (1986).

\bibitem{sobota} J. A. Sobota, D. Tanaskovic, and V. Dobrosavljevic,
  Phys. Rev. B {\bf 76}, 245106 (2007).

\bibitem{white} R. M. White, {\it Quantum Theory of Magnetism}
  (McGraw-Hill, New York, 1970).

\bibitem{alexander} A. Wei\ss e, G. Wellein, A. Alvermann, and
  G. Fehske, Rev. Mod. Phys. {\bf 78}, 275 (2006).

\bibitem{amini} M. Amini, S. A. Jafari, and F. Shahbazi,
  Euro. Phys. Lett. {\bf 87} 37002 (2009).

\bibitem{lherbier} A. Lherbier, B. Biel, Y. M. Niquet, and S. Roche,
  Phys.  Rev. Lett. {\bf 100}, 036803 (2008).

\bibitem{neto} A. H. Castro Neto {\it et al.}, Rev. Mod. Phy. {\bf
  81}, 109 (2009).

\end{thebibliography}
\end{document}